\documentclass{IOS-Book-Article}

\usepackage{times}
\normalfont
\usepackage[T1]{fontenc}
\usepackage{amssymb}
\usepackage{multirow}
\usepackage{graphicx}
\usepackage{palatino}
\usepackage{url}
\usepackage{cite}
\usepackage{multicol}
\usepackage{listings}
\usepackage{color}
\usepackage{subfigure}
\usepackage{ifpdf}
\usepackage{epsfig}
\usepackage{color}
\newcommand{\Cpp}{C\kern-0.05em\texttt{+\kern-0.03em+}}

\long\gdef\comment#1#2{}

\newcommand{\code}[1]{\lstinline[basicstyle=\sffamily]{#1}}

\newcommand{\ConceptCpp}{ConceptC\kern-0.05em\texttt{+\kern-0.03em+}}
\newcommand{\MPIpp}{MPI\kern-0.05em\texttt{+\kern-0.03em+}}
\newcommand{\Cilkpp}{Cilk\kern-0.05em\texttt{+\kern-0.03em+}}
\newcommand{\Charmpp}{Charm\kern-0.05em\texttt{+\kern-0.03em+}}

\renewcommand{\emph}{\textit}

\lstdefinestyle{basic}{showstringspaces=false,columns=fullflexible,language=C++,escapechar=@,xleftmargin=1pc,%
basicstyle=\small\sffamily,
commentstyle=\mdseries,
moredelim=**[is][\color{white}]{~}{~},
morekeywords={concept,model,require,where,reduction,cilk,spawn,sync,omp,pragma,task,taskwait},
literate={->}{{$\rightarrow\;$}}1 {<-}{{$\leftarrow\;$}}1 {=>}{{$\Rightarrow\;$}}1,
}

\makeatletter

\begin{document}
\begin{frontmatter}                           

\title{Extending Task Parallelism For Frequent Pattern Mining}

\author[A]{\fnms{Prabhanjan Kambadur}},
\author[B]{\fnms{Amol Ghoting}},
\author[B]{\fnms{Anshul Gupta}} and
\author[A]{\fnms{Andrew Lumsdaine}}

\runningauthor{Kambadur et al.}
\address[A]{Open Systems Lab, Indiana University, Bloomington, IN - 47408}
\address[B]{IBM T J Watson Research Center, Yorktown Heights, NY - 10598}

\begin{abstract}
Algorithms for frequent pattern mining, a popular informatics application,
have unique requirements that are not met by any of the existing parallel 
tools. In particular, such applications operate on extremely large data
sets and have irregular memory access patterns. For efficient parallelization
of such applications, it is necessary to support dynamic load balancing
along with scheduling mechanisms that allow users to exploit data locality.  
Given these requirements, task parallelism is the most promising of the
available parallel programming models.  However, existing solutions for task
parallelism schedule tasks implicitly and hence, custom scheduling policies
that can exploit data locality cannot be easily employed.
In this paper we demonstrate and characterize the speedup obtained in a
frequent pattern mining application using a custom clustered scheduling
policy in place of the popular Cilk-style policy.  We present PFunc, a novel
task parallel library whose customizable task scheduling and task priorities
facilitated the implementation of our clustered scheduling policy.
\end{abstract}

\begin{keyword}
task parallelism, frequent pattern mining, data locality.
\end{keyword}

\end{frontmatter}

\section*{Introduction}
Algorithms for frequent pattern mining (FPM), a popular informatics
application, exhibit certain distinct characteristics that distinguish them
from traditional high-performance computing applications. Typically, FPM
applications operate on large, irregular and dynamic data sets.  The data (and
corresponding computations) cannot be partitioned \textit{a priori}, making
these applications extremely sensitive to load balancing and scheduling.
Memory access patterns are often data-dependent, requiring one data object's
location in memory to be resolved before the next can be fetched. Finally,
safe parallelization of FPM applications requires fine-grained synchronization.
For all of these reasons, popular parallel programming models such as the data
parallel and the single process multiple data (SPMD) models are not well-suited
to FPM applications.
Of the variety of alternative parallel programming models available, task
parallelism is the most promising when it comes to meeting the challenges posed
by FPM applications. The task parallel programming model is sufficiently
high-level and general purpose to be able to parallelize both regular and
irregular applications. However, existing solutions for task parallelism have
shortcomings that prevent efficient parallelization of FPM applications.
Specifically, FPM applications require custom task scheduling policies that can
exploit data locality between tasks that are not always related by the
parent-child relationship. In most existing solutions for task parallelism, not
only are tasks scheduled transparently from the users, but also, data locality
between tasks is exploited only if there is a parent-child relationship between
those two tasks. 

In this paper, we demonstrate the speedup obtained in a task parallel
Apriori-based FPM implementation by switching from the popular
Cilk-style~\cite{Blumofe94} task scheduling policy to a customized
``clustered'' task scheduling policy. Furthermore, we collect various hardware
metrics to characterize the factors that resulted in the speedup.  To implement
these two (Cilk-style and clustered) scheduling policies, we use PFunc, a novel
library-based implementation of task parallelism, that allows users to
customize parameters such as task scheduling policy and task priority. Unlike
most other parallelization tools, PFunc provides a natural interface that
enables facile implementation of our clustered scheduling policy.  Through this
study, we highlight the need to incorporate support for efficient
parallelization of FPM applications in the task parallel model.

\section{Background}
\label{sec:back}
Murphy and Kogge~\cite{Kogge:2007} demonstrated the differences in memory
access patterns of informatics applications when compared to those of
traditional scientific computing applications.  Berry et al.~\cite{Berry:2007}
and Lumsdaine et al.~\cite{Lumsdaine:2007} have shown that current techniques
applied to high performance computing are inadequate for informatics
applications.
Since FPM was introduced as a relevant problem in informatics, its memory
characteristics and parallelization have been extensively
researched~\cite{Agrawal:1994, Agrawal:1996, Sam:2000,
Parthasarathy:2001,Buehrer:2006,Nijsen:2004,Wang:2004,Han:2000, Ghoting:2007,
Kim:1998}.  Zaki and Parathasarathy~\cite{Zaki:1997} were the first to explore
the idea of clustering to aid in fast discovery of frequent patterns.  

Many solutions have been implemented to facilitate dynamic task parallelism.
Fortran M~\cite{Foster97}, Cilk~\cite{FrigoLeRa98} and OpenMP
3.0~\cite{kn:omp_30} implement task parallelism as extensions to stock
programming languages. Other solutions such as Intel's Threading Building
Blocks (TBB)~\cite{kn:tbb}, Microsoft's Parallel Patterns Library (PPL) and
Task Parallel Library (TPL), and Java Concurrency Utilities are library-based.
All of the three HPCS languages (Chapel~\cite{Chamberlain:2007p1040},
Fortress~\cite{fortress} and X10~\cite{Charles:2005p1232}) offer task
parallelism as language features. Scheduling of tasks has received wide
attention in the programming community. Cilk's \textit{depth-first
work}~\cite{Blumofe94} model, the X10 Work Stealing framework's (XWS)
\textit{breadth-first}~\cite{Cong08} model and Guo et al.'s \textit{hybrid}
model~\cite{Sarkar09} are notable examples of work stealing schedulers.  Each
of the three above mentioned scheduling policies exploit data locality under
different circumstances. For example, Cilk's scheduling policy exploits data
locality only when the applications are deeply nested. 
All of the task parallel solutions discussed above schedule tasks implicitly,
thereby disallowing customization of task scheduling policies. 
In this paper, we demonstrate the importance of customizing the task scheduling
policy in parallelization tools when parallelizing FPM applications.  We also
describe the features of PFunc that enable FPM applications to employ custom
scheduling strategies that help them outperform other implementations that
employ default scheduling policies provided by existing solutions for task
parallelism.

\section{Problem Description}
\label{sec:problem}
In this section, we describe FPM and comment on important aspects of its
implementation that influence performance.  Briefly, the problem description is
as follows: Let $I=\{i_{1},i_{2},..,i_{n}\}$ be a set of $n$ items, and let
$D=\{T_{1},T_{2},..,T_{m}\}$  be a set of $m$ transactions, where each
transaction $T_{i}$ is a subset of $I$. An itemset $i \subseteq{} I$ of size
$k$ is known as a $k$-itemset. The support (that is, frequency) of $i$ is
$\sum{}_{j=1}^{m} (1:i\subseteq{}T_{j})$, or informally speaking, the number of
transactions in $D$ that have $i$ as a subset. The FPM problem is to find all
$i \in{} D$ that have support greater than a user supplied minimum value.  
For our FPM implementation, we choose the \textit{Apriori}~\cite{
Agrawal:1994:Apriori} algorithm.  Due to its efficiency, robustness and
guaranteed main memory footprint, the Apriori algorithm is widely used in FPM
implementations including those in commercial products such as IBM's InfoSphere
Warehouse~\cite{zaki1999}.   Apriori traverses the itemset search space in
breadth-first order.  Its efficiency stems from its use of the anti-monotone
property: If a size \textit{k}-itemset is not frequent, then any size
\textit{(k+1)}-itemset containing it will not be frequent. The algorithm
first finds all frequent \textit{1}-items in the data set, and then
iteratively finds all frequent \textit{k}-itemsets using the frequent
\textit{(k-1)}-itemsets discovered previously. For example, let A, B, C and D
be individual items (\textit{1}-itemsets) that are frequent in a transaction
database. Then, for stage 2, AB, AC, AD, BC, BD, and CD are the candidate
\textit{2}-itemsets. If at stage 2, after counting, \textit{2}-itemsets AB,
AC and AD were found to be frequent, then ABC, ABD, and ACD are the
candidates for stage 3.  The frequency of a candidate \textit{k}-itemset is
counted by performing a join ($\Join{}$) operation on the transaction-ID
lists of each individual item in that particular itemset. In our task
parallel implementation of the Apriori algorithm, the counting operations
required for each \textit{k}-itemset are executed as a separate task. 

\paragraph{Requirements for efficient parallelization:}
The Apriori algorithm is highly dependent on memory reuse for its
performance~\cite{Ghoting:2007}. As tasks mine for itemsets, they access
overlapping memory regions as many itemsets share transaction-ID lists.  For
example, if we were mining for \textit{2}-itemsets AB, AC, and AD, then the
transaction-ID list of A would be common in the mining operations. The greater
the overlap of items in the itemsets, the greater the potential for memory
reuse between their respective tasks. Exploiting such inter-task data
localities through locality-aware scheduling of tasks is the key for efficient
parallel execution of Apriori-based FPM applications.

\paragraph{Shortcomings of current solutions:}
Existing task parallel solutions schedule tasks implicitly using different
flavors of the Cilk-style work stealing~\cite{Blumofe94} task scheduler and
hence, do not support custom task scheduling policies that can exploit data
localities between user-specified tasks. Furthermore, in the
Cilk-style scheduling policy, there is little or no
contention when a thread executes tasks that are on its own task queue, but stealing
a task from another thread's task queue is an expensive operation.  Cilk-style
work stealing benefits applications that are deeply nested (that is, recursive) in
nature. Nested tasks, by definition, spawn other tasks. Therefore, when a
thread steals a nested task, it implicitly gains all the tasks that will be
generated by the stolen task--thereby minimizing task stealing.  However, in
the Apriori algorithm, tasks are non-nested as we traverse the search space in
breadth-first order. Hence, once a thread runs out of work, it must steal tasks
repeatedly from other \textit{victim} threads' task queues in order to keep
itself busy.  Such work stealing is detrimental to Apriori-based FPM
implementations' performance both because of the increased contention on victim
threads' task queues and the lack of data locality among stolen tasks.

\section{PFunc: A new tool for task parallelism}
\label{sec:pfunc}

PFunc~\cite{kambadur09:pfunc} is a lightweight and portable task parallel
library for C and \Cpp{} users, which has been designed using the generic
programming paradigm~\cite{garcia05:_extended_comparing05} to overcome some of
the shortcomings of existing solutions for task parallelism.  Due to space
constraints, we describe only those features of PFunc that affect Apriori-based
FPM implementations.
PFunc allows users to choose the task scheduling policy at compile time. Users
can either choose from built-in scheduling policies such as Cilk-style,
first-in first-out (FIFO), last-in first-out (LIFO) and priority-based, or
they can supply their own scheduling policy. All scheduling policies are
``models'' of the \code{scheduler} ``concept''. This generic design enforces a
uniform interface across the different scheduling policies and enables compile
time plug-and-play capabilities with no runtime penalty.
By default, PFunc follows the work stealing model in which each thread has its
own task queue and tasks are enqueued on the task queue of the thread that
spawned the task.  Users have the option to override this default setting at
runtime and place tasks onto a particular thread's task queue (that is, change
the task's \textit{affinity}).
PFunc also provides \textit{task attributes} that can be attached to each
individual task when spawning the task. Task attributes, such as task
priority, are an important tool in the implementation of many scheduling
policies. For example, when using priority-based scheduling, users specify a
task's priority using the task attribute mechanism. Like the scheduler, task
attributes can be customized to suit the task scheduling policy. 
To enable hardware profiling, PFunc is fully integrated with the Performance
Application Programming Interface (PAPI)~\cite{papi}.

\section{Clustered task scheduling}
\label{sec:cluster}
To efficiently parallelize our Apriori-based FPM implementation using the task
parallel programming model, we have designed a clustered scheduling policy.
In this policy, tasks are clustered together such that there is a greater
likelihood of memory reuse between the clustered tasks (for example, tasks
mining for \textit{3}-itemsets ABC and ABD) as they are executed by the same
thread.  Furthermore, we employ a custom task stealing policy in which clusters
of tasks are stolen instead of a single task.  Such clustered stealing not
only reduces contention on the thread-local task queues by minimizing the
number of required steals, but also maintains data locality between the stolen
tasks.

\paragraph{Implementation:} In our FPM implementation, each task performs the
mining operations required for one \textit{k}-itemset. Each \textit{k}-itemset
is stored as a \textit{sorted} set of individual items. 
To cluster tasks that have significant memory overlap, we use a common prefix
of length (k-1).  For example, the \textit{3}-itemsets ABC and ABD share the
2-prefix AB and hence, their respective tasks are clustered together. For
efficient execution of our FPM application, it is necessary to ensure that such
clusters of tasks have a high likelihood of being executed by the same thread.
To this end, we use a hash table (\code{std::hash_map}) as the task queue for
each thread.  This is a departure from the Cilk model~\cite{FrigoLeRa98}, in
which each thread's task queue is a deque. With the hash table structure in
place, clustering of tasks is achieved by placing all \textit{k}-itemsets that
have a common (k-1) prefix into the same bucket. To achieve such clustering, we
used the following hash function to compute the hash of a \textit{k}-itemset.
First, the hash of each of its first (k-1) items is computed using \Cpp{}
Standard Library's~\cite{stepa.lee-1994:the.s:TR} \code{std::hash} function
object. Then, these individual hashes are XOR'ed together to produce one final
hash.  For example, the hash for the \textit{3}-itemset ABC is computed by
XOR-ing the results of applying \code{std::hash} to A and B separately. The
hashes thus computed for the \textit{3}-itemsets ABC and ABD are equal
and hence, their respective tasks are placed in the same bucket.
Each thread executes the tasks placed in its task queue (that is, hash table)
by iterating through its task queue's buckets starting from the first non-empty
bucket. 
When a thread runs out of tasks on its own task queue, it randomly selects a
victim thread to steal tasks. Then, it steals the first non-empty bucket from
the victim thread's task queue. This stealing policy is better equipped than
the Cilk-style stealing policy to avoid repeated stealing in our FPM
implementation as potentially, more than one task can be stolen during each
steal.  Furthermore, by stealing an entire bucket of tasks, data locality
between stolen tasks is preserved.

\paragraph{Integrating with PFunc:} Realizing our customized clustered task
scheduling policy in PFunc is achieved in two steps. 
First, the entire scheduling policy is implemented to model PFunc's 
\code{scheduler} concept. This enables us to pick our clustered scheduling
policy as the task scheduling policy of choice at compile time. 
Second, we customize PFunc's task attributes (again, at compile time) such that
we can attach a \textit{reference} to the \textit{k}-itemset that needs to be
mined as the respective task's priority. When a task is spawned, our hash
function uses its priority (that is, the associated \textit{k}-itemset's
reference) to determine the appropriate bucket (that is, cluster) for this task
in the spawning thread's task queue.

\section{Results}
\label{sec:results}

\begin{figure}[t]
\includegraphics[width=0.6\textwidth]{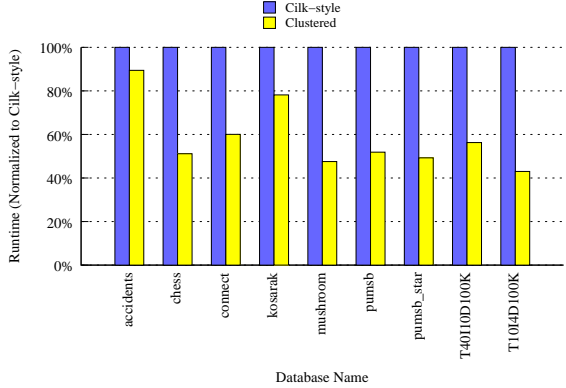}
\caption{Graph showing the normalized runtimes of different datasets when using
the Cilk-style and Clustered task scheduling policies in our Apriori-based
FPM implementation with 8 threads. Support (that is, frequency) for each of
the datasets is given in Table~\ref{tbl:fim8}.}
\label{fig:fim8}
\end{figure}

In this section, we present the results of running our Apriori-based FPM
implementation using the Cilk-style and the clustered scheduling policies.
PFunc was used for parallelization of our FPM implementation and the scheduling
policy for a particular run was chosen at compile time. Minimal code
modification was required to attach the \textit{k}-itemset reference as task
priority when spawning the tasks for the clustered scheduling policy.
We ran our experiments on a four socket, quad-core (total 16 cores) AMD 8356
processor running Linux Kernel 2.6.24.  For compilation, we used GCC 4.3.2
with: ``-O3 -fomit-frame-pointer -funroll-loops''. To collect hardware
metrics, we used PFunc's integration with PAPI.  The benchmarks were run on
data sets from the FIMI repository~\cite{bart:2004}.

\begin{table*}[t]
\centering
\begin{tabular}{|c|c|c|c|c|c|c|c|}  \hline

\multirow{2}{*}{Dataset} & 
\multirow{2}{*}{Support} & 
\multicolumn{2}{|c|}{IPC} &
\multicolumn{2}{|c|}{DTLB L1M/L2H} &
\multicolumn{2}{|c|}{DTLB L1M/L2M} \\ \cline{3-8}
& & Cilk & Cluster & 
Cilk & Cluster & 
Cilk & Cluster \\ 
\hline
accidents & 0.25 &  0.595689 & 0.603959 &
0.000048 & 0.000046 & 0.000161 & 0.000110\\ \hline
chess & 0.6 & 0.560538 & 0.668965 & 
0.000797 & 0.000242 & 0.001006 & 0.000032\\ \hline
connect & 0.8 & 0.543099 & 0.809308 & 
0.000249 & 0.000112 & 0.001204 & 0.000141\\ \hline
kosarak & 0.0013 & 0.692103 & 0.717599 & 
0.000400 & 0.000185 & 0.000659 & 0.000123\\ \hline
pumsb & 0.75 & 0.494539 & 0.719072 & 
0.000230 & 0.000114 & 0.001276 & 0.000126\\ \hline
pumsb\_star & 0.3 & 0.527659 & 0.698358 & 
0.000315 & 0.000145 & 0.001082 & 0.000113\\ \hline
mushroom & 0.10 & 0.570390 & 0.705003 & 
0.000477 &  0.000267 & 0.000950 &0.000022\\ \hline
T40I10D100K & 0.005 & 0.627272 & 0.727288 &
0.000368 & 0.000305 & 0.000900 & 0.000021\\ \hline
T10I4D100K & 0.00006 & 0.555330 & 0.716282 &
0.000218 & 0.000144 & 0.000876 & 0.000044 \\ \hline
\end{tabular}
\vspace{+5pt}
\caption{Instructions-per-cycle (IPC), L1 data TLB misses (L1M/L2H) and L2 data
TLB misses (L1M/L2M) when using the Cilk-style and Clustered task scheduling
policies in our Apriori-based FPM implementation with 8 threads for different
datasets from the FIMI repository~\cite{bart:2004}.}
\label{tbl:fim8}
\end{table*}

Figure~\ref{fig:fim8} depicts the runtimes of our FPM implementation for both
the Cilk-style and the clustered task scheduling policies. Runtimes were
recorded with hardware profiling turned off and were averaged after $5$ runs.
The clustered scheduling policy runs significantly faster (more than $50\%$)
for most of the data sets, with the \code{accidents.dat} data set being the
only exception. To test our hypothesis that our clustered scheduling policy
exploits data locality better than the Cilk-style policy, we collected
various hardware metrics.  Table~\ref{tbl:fim8} summarizes some of the
important metrics. Our clustered scheduling policy delivers more instructions
per clock cycle than the Cilk-style scheduling policy for all data sets.
Also, our clustered scheduling policy incurs far fewer L2 data TLB misses
than the Cilk-style scheduling policy. This is because the benefits of
clustering tasks that have a significant memory overlap outweigh the
additional operational costs incurred due to using hash tables in our
clustered scheduling policy.
When results from Figure~\ref{fig:fim8} and Table~\ref{tbl:fim8} are taken
together, we can conclude that the clustered scheduling policy exploits
data locality better than the Cilk-style scheduling policy for our
Apriori-based FPM implementation.

\section{Conclusion and Future Work}
We have demonstrated that our custom clustered scheduling policy performs
better than the popular Cilk-style scheduling policy for an Apriori-based FPM
implementation. This was made possible by PFunc, which provides better support
for parallelization of FPM applications than existing solutions for task
parallelism by allowing facile customization of task scheduling policy and
task attributes. An interesting topic for future research is to implement a
dynamic task scheduling policy that utilizes a multi-dimensional index
structure as task queues.  This would enable a thread to dynamically pick the
``nearest-neighbor'' of the previously executed task as its next task to
execute. 

\paragraph{\textbf{Acknowledgments:}}
We thank Melanie Dybvig for her help in improving the quality of this paper.
Our work was supported by IBM, King Abdullah University of Science and
Technology (KAUST), National Science Foundation grants EIA-0202048 and
CCF-0541335, and a grant from the Lilly Endowment.

\bibliographystyle{abbrv}
\bibliography{refs}

\begin{thebibliography}{10}

\bibitem{papi}
{\em {Performance Application Programming Interface}}.
\newblock ICL, Knoxville, TN.

\bibitem{bart:2004}
{\em {Datasets of the workshops on Frequent Itemset Mining Implementations
  (FIMI)}}.
\newblock University of Helsinki, 2004.

\bibitem{Agrawal:1994}
R.~Agrawal, T.~Imieli\'{n}ski, and A.~Swami.
\newblock {Mining association rules between sets of items in large databases}.
\newblock In {\em SIGMOD '93: Proceedings of the 1993 ACM SIGMOD international
  conference on Management of data}, pages 207--216, New York, NY, USA, 1993.
  ACM.

\bibitem{Agrawal:1996}
R.~Agrawal and J.~C. Shafer.
\newblock {Parallel Mining of Association Rules}.
\newblock {\em IEEE Trans. on Knowl. and Data Eng.}, 8(6):962--969, 1996.

\bibitem{Agrawal:1994:Apriori}
R.~Agrawal and R.~Srikant.
\newblock {Fast Algorithms for Mining Association Rules in Large Databases}.
\newblock In {\em VLDB '94: Proceedings of the 20th International Conference on
  Very Large Data Bases}, pages 487--499, San Francisco, CA, USA, 1994. Morgan
  Kaufmann Publishers Inc.

\bibitem{fortress}
E.~Allen, D.~Chase, J.~Hallett, V.~Luchangco, J.-W. Maessen, S.~Ryu, G.~L.~S.
  Jr., and S.~Tobin-Hochstadt.
\newblock {The Fortress Language Specification, Version 1.0}.
\newblock Technical report, Sun Microsystems, Inc., 2008.

\bibitem{Berry:2007}
J.~W. Berry, B.~Hendrickson, S.~Kahan, and P.~Konecny.
\newblock {Software and Algorithms for Graph Queries on Multithreaded
  Architectures}.
\newblock {\em Parallel and Distributed Processing Symposium, 2007. IPDPS 2007.
  IEEE International}, 2007.

\bibitem{Blumofe94}
R.~D. Blumofe and C.~E. Leiserson.
\newblock Scheduling {M}ultithreaded {C}omputations by {W}ork {S}tealing.
\newblock In {\em Proceedings of the 35th Annual Symposium on Foundations of
  Computer Science (FOCS}, pages 356--368, 1994.

\bibitem{Buehrer:2006}
G.~Buehrer, S.~Parthasarathy, and Y.-K. Chen.
\newblock {Adaptive Parallel Graph Mining for CMP Architectures}.
\newblock {\em Data Mining, IEEE International Conference on}, 0:97--106, 2006.

\bibitem{Chamberlain:2007p1040}
B.~L. Chamberlain, D.~Callahan, and H.~P. Zima.
\newblock {Parallel Programmability and the Chapel Language}.
\newblock {\em International Journal of High Performance Computing
  Applications}, Jan 2007.

\bibitem{Charles:2005p1232}
P.~Charles, C.~Grothoff, V.~Saraswat, C.~Donawa, A.~Kielstra, K.~Ebcioglu,
  C.~von Praun, and V.~Sarkar.
\newblock X10: an object-oriented approach to non-uniform cluster computing.
\newblock In {\em OOPSLA '05: Proceedings of the 20th annual ACM SIGPLAN
  conference on Object-oriented programming, systems, languages, and
  applications}, pages 519--538, New York, NY, USA, 2005. ACM.

\bibitem{Cong08}
G.~Cong, S.~Kodali, S.~Krishnamoorthy, D.~Lea, V.~Saraswat, and T.~Wen.
\newblock Solving {L}arge, {I}rregular {G}raph {P}roblems {U}sing {A}daptive
  {W}ork-{S}tealing.
\newblock In {\em ICPP '08: Proceedings of the 2008 37th International
  Conference on Parallel Processing}, pages 536--545, Washington, DC, USA,
  2008. IEEE Computer Society.

\bibitem{Foster97}
I.~Foster, D.~R. Kohr, Jr., R.~Krishnaiyer, and A.~Choudhary.
\newblock {A Library-based Approach to Task Parallelism in a Data-parallel
  Language}.
\newblock {\em J. Parallel Distrib. Comput.}, 45(2):148--158, 1997.

\bibitem{FrigoLeRa98}
M.~Frigo, C.~E. Leiserson, and K.~H. Randall.
\newblock The implementation of the {C}ilk-5 multithreaded language.
\newblock In {\em Proceedings of the ACM SIGPLAN '98 Conference on Programming
  Language Design and Implementation}, pages 212--223, Montreal, Quebec,
  Canada, June 1998.
\newblock Proceedings published in ACM SIGPLAN Notices, Vol. 33, No. 5, May,
  1998.

\bibitem{garcia05:_extended_comparing05}
R.~Garcia, J.~J\"arvi, A.~Lumsdaine, J.~Siek, and J.~Willcock.
\newblock An extended comparative study of language support for generic
  programming.
\newblock {\em Journal of Functional Programming}, 2005.

\bibitem{Ghoting:2007}
A.~Ghoting, G.~Buehrer, S.~Parthasarathy, D.~Kim, A.~Nguyen, Y.-K. Chen, and
  P.~Dubey.
\newblock Cache-conscious frequent pattern mining on modern and emerging
  processors.
\newblock {\em The VLDB Journal}, 16(1):77--96, 2007.

\bibitem{Sarkar09}
Y.~Guo, R.~Barik, R.~Raman, and V.~Sarkar.
\newblock Work-{F}irst and {H}elp-{F}irst {S}cheduling {P}olicies for
  {A}sync-{F}inish {T}ask {P}arallelism.
\newblock In {\em Proceedings of the 23rd IEEE International Parallel and
  Distributed Processing Symposium}, May 2009.

\bibitem{Sam:2000}
E.-H.~S. Han, G.~Karypis, and V.~Kumar.
\newblock {Scalable Parallel Data Mining for Association Rules}.
\newblock {\em IEEE Trans. on Knowl. and Data Eng.}, 12(3):337--352, 2000.

\bibitem{Han:2000}
J.~Han, J.~Pei, and Y.~Yin.
\newblock {Mining frequent patterns without candidate generation}.
\newblock In {\em SIGMOD '00: Proceedings of the 2000 ACM SIGMOD international
  conference on Management of data}, pages 1--12, New York, NY, USA, 2000. ACM.

\bibitem{kambadur09:pfunc}
P.~Kambadur, A.~Gupta, A.~Ghoting, H.~Avron, and A.~Lumsdaine.
\newblock {PFunc: Modern Task Parallelism For Modern High Performance
  Computing}.
\newblock In {\em SC '09: Proceedings of the 2008 ACM/IEEE conference on
  Supercomputing}, Portland, Oregon, November 2009.

\bibitem{Kim:1998}
J.-S. Kim, X.~Qin, and Y.~Hsu.
\newblock Memory characterization of a parallel data mining workload.
\newblock In {\em WWC '98: Proceedings of the Workload Characterization:
  Methodology and Case Studies}, page~60, Washington, DC, USA, 1998. IEEE
  Computer Society.

\bibitem{Lumsdaine:2007}
A.~Lumsdaine, D.~Gregor, B.~Hendrickson, and J.~W. Berry.
\newblock Challenges in parallel graph processing.
\newblock {\em Parallel Processing Letters}, 17(1):5--20, 2007.

\bibitem{Kogge:2007}
R.~C. Murphy and P.~M. Kogge.
\newblock {On the Memory Access Patterns of Supercomputer Applications:
  Benchmark Selection and Its Implications}.
\newblock {\em IEEE Trans. Comput.}, 56(7):937--945, 2007.

\bibitem{Nijsen:2004}
S.~Nijssen and J.~N. Kok.
\newblock {A quickstart in frequent structure mining can make a difference}.
\newblock In {\em KDD '04: Proceedings of the tenth ACM SIGKDD international
  conference on Knowledge discovery and data mining}, pages 647--652, New York,
  NY, USA, 2004. ACM.

\bibitem{kn:omp_30}
{O}penMP {A}rchitecture~{R}eview {B}oard.
\newblock {\em OpenMP Application Program Interface, v3.0}.
\newblock May 2008.

\bibitem{Parthasarathy:2001}
S.~Parthasarathy, M.~J. Zaki, M.~Ogihara, and W.~Li.
\newblock Parallel data mining for association rules on shared memory systems.
\newblock {\em Knowl. Inf. Syst.}, 3(1):1--29, 2001.

\bibitem{kn:tbb}
J.~Reinders.
\newblock {\em Intel Threading Building Blocks}.
\newblock O'Reilly, 2007.

\bibitem{stepa.lee-1994:the.s:TR}
A.~A. Stepanov and M.~Lee.
\newblock {The Standard Template Library}.
\newblock Technical Report X3J16/94-0095, WG21/N0482, ISO Programming Language
  C++ Project, May 1994.

\bibitem{Wang:2004}
C.~Wang and S.~Parthasarathy.
\newblock {Parallel algorithms for mining frequent structural motifs in
  scientific data}.
\newblock In {\em ICS '04: Proceedings of the 18th annual international
  conference on Supercomputing}, pages 31--40, New York, NY, USA, 2004. ACM.

\bibitem{zaki1999}
M.~J. Zaki.
\newblock {Parallel and Distributed Association Mining: A Survey}.
\newblock {\em IEEE Concurrency}, 7(4):14--25, 1999.

\bibitem{Zaki:1997}
M.~J. Zaki, S.~Parthasarathy, M.~Ogihara, and W.~Li.
\newblock {New Algorithms for Fast Discovery of Association Rules}.
\newblock Technical report, Rochester, NY, USA, 1997.

\end{thebibliography}

\end{document}